\documentclass[12pt]{article}
\usepackage{latexsym}
\usepackage{amsmath}
\usepackage{amssymb}
\usepackage{epsfig,graphics}
\usepackage{graphicx}
\usepackage{booktabs}
\usepackage{multirow}
\usepackage{caption}
\usepackage{subcaption}
\usepackage{tikz}
\usepackage{fixmath}
\usetikzlibrary{matrix,snakes,arrows,shapes,decorations.pathmorphing,decorations.markings,calc}

\newcommand{\be}{\begin{equation}}
\newcommand{\ee}{\end{equation}}

\begin{document}

\begin{titlepage}

\vspace*{0.6in}
 
\begin{center}
{\large\bf Spinorial flux tubes in SO(N) gauge theories in 2+1 dimensions}\\
\vspace*{0.75in}
{Michael Teper \\
\vspace*{.25in}
Rudolf Peierls Centre for Theoretical Physics, University of Oxford,\\
1 Keble Road, Oxford OX1 3NP, UK\\
\centerline{and}
All Souls College, University of Oxford,\\
High Street, Oxford OX1 4AL, UK}
\end{center}

\vspace*{0.4in}

\begin{center}
{\bf Abstract}
\end{center}

We investigate whether one can observe in $SO(3)$ and $SO(4)$ (lattice) gauge theories
the presence of  spinorial flux tubes, i.e. ones that correspond to the
fundamental representation of $SU(2)$; and similarly for $SO(6)$ and $SU(4)$.
We do so by calculating the finite volume dependence of the $J^p=2^+$ glueball
in $2+1$ dimensions, using lattice simulations. We show how this
provides strong evidence that these $SO(N)$ gauge theories contain states that
are composed of (conjugate) pairs of winding spinorial flux tubes, i.e. ones that are
in the (anti)fundamental of the corresponding $SU(N^\prime)$ gauge theories.
Moreover, these two flux tubes can be arbitrarily far apart. This is so
despite the fact that the fields that are available in the $SO(N)$ lattice field
theories do not appear to allow us to construct operators that project onto
single spinorial flux tubes.

\vspace*{0.95in}

\leftline{{\it E-mail:} mike.teper@physics.ox.ac.uk}

\end{titlepage}

\setcounter{page}{1}
\newpage
\pagestyle{plain}

\tableofcontents

\section{Introduction}
\label{section_intro}

There are pairs of $SU(N)$ and $SO(N^\prime)$ gauge theories that share the same
Lie algebra. These pairs are $SO(3)$ and $SU(2)$, $SO(4)$ and $SU(2)\times SU(2)$,
and $SO(6)$ and $SU(4)$. It is interesting to ask how, if at all, the differing
global properties of the groups in each pair affect the dynamics. This question
has been addressed using lattice methods in 
\cite{RLMT_GK,MT_GK,RLMT_Tc}
for gauge theories in 2+1 dimensions. It turns out
\cite{RLMT_GK,MT_GK}
that those glueball masses that have been reliably calculated, in particular
the $J^P=0^+$ and $J^P=2^+$ ground states, are indeed consistent within each
pair, as are string tensions and couplings.
So also is the deconfining temperature, $T_c$, for $SO(4)$ compared to
$SU(2)\times SU(2)$ and for $SO(6)$ compared to $SU(4)$
\cite{RLMT_Tc}.
(There is no calculation of $T_c$ for $SO(3)$.) The lightest excited glueball
states, although not so reliably calculated, also appear to be consistent
\cite{MT_GK}.
All this suggests quite strongly that the low-lying
non-perturbative physics is the same within each of these pairs of gauge theories.

This raises some interesting questions. The comparison between the pairs of theories
can only be made in their common sector of states. So consider for example
a periodic space-time and a fundamental flux tube that winds around one of the
spatial tori. In, say, $SO(6)$ such a flux tube with flux in the fundamental
representation corresponds to a flux tube in $SU(4)$ with flux in the $k=2A$
representation (the antisymmetric piece of the product of two fundamentals).
And indeed one finds that the corresponding string tensions are equal when
expressed in units of the coupling or the mass gap
\cite{RLMT_GK,MT_GK}.
On the other hand the fundamental flux tube of $SU(4)$ corresponds to the
spinorial of $SO(6)$ and this cannot be constructed from tensor products of
the  $SO(6)$ fundamental representation. That is to say, we cannot construct an
operator using our $SO(6)$ lattice gauge fields that projects onto such a
spinorial flux tube and therefore one might expect such a flux tube not to exist
in the spectrum of the $SO(6)$ theory. However a state consisting of two winding
flux tubes, one spinorial and one conjugate-spinorial, can have vacuum quantum
numbers and should be accessible to our operators. If these winding flux tubes
are sufficiently far apart from each other then they will not interact and
this appears paradoxical if the theory does not possess states consisting
of  single spinorial flux tubes.

In this paper this is the issue we address.
Since the $SO(N)$ gauge theory does not contain operators that project onto a
single spinorial flux tube, our strategy  will have to be indirect.
We use the fact that in an $SU(N)$ gauge theory on sufficiently small volumes the
lightest glueballs are composed of a pair of (conjugate) winding flux tubes
in the fundamental
representation. This can occur because a state consisting of a winding flux tube and
its conjugate will have a non-zero overlap onto the usual contractible loops that
form a basis for glueball operators. (In contrast to a single winding flux tube
that has zero overlap by the standard centre symmetry argument.)
The reason it actually does occur is that  the energy of a winding flux
tube decreases with its length (roughly linearly) and so as the lattice volume
decreases a conjugate pair of such flux tubes should eventually become
lighter than the `normal' glueball state, and will then become the ground state
glueball. The signal for this happening is thus the sudden onset of a rapid
decrease in the
glueball ground state energy as the volume decreases. If we find exactly the
same finite volume behaviour in, say, $SO(6)$ as in $SU(4)$, then we can infer that
the $SO(6)$ theory possesses states composed of a pair of (conjugate) flux tubes
in the spinorial representation of $SO(6)$, i.e. ones that are in the fundamental of
$SU(4)$. And similarly for $SO(3)$ and $SU(2)$ and also for $SO(4)$ and
$SU(2)\times SU(2)$. This is indeed what we shall find. This is not such a
surprise if one believes that the finite volume spectra of such pairs of $SO(N)$
and $SU(N^\prime)$ theories should be identical --  but of course this need not
be the case, especially since the effect of the differing global properties of the
groups might well be enhanced on small volumes.

There is a corresponding puzzle in the context of the glueball spectrum which arises
because the $SO(N)$ group elements are real so naively the fields available in the
theory do not allow us to write an operator with $C=-$. Thus, for example, the $C=-$
states of $SU(4)$ would appear to have no counterpart in $SO(6)$, just like the
fundamental flux tubes of  $SU(4)$ discussed above. (These are not unrelated: after
all contractible closed loops of fundamental flux can provide a basis for $C=-$
glueballs in  $SU(4)$.) However a pair of $C=-$ glueballs will have $C=+$
and so one can ask whether such `scattering' states appear in the $SO(N)$ theory.
Indeed in $SU(4)$ an excited  $J^{PC}=0^{++}$ glueball that has a mass larger
than about 3 times the mass gap has enough energy to decay into two $0^{--}$ glueballs
\cite{AAMT_SUN}
and will do so with some non-zero decay width. So if the low-lying $SU(4)$ and $SO(6)$ 
$J^{PC}=0^{++}$ spectra are identical, the decay widths will be the same and such
decay products of two $0^{--}$ glueballs will be present in the $SO(6)$ gauge theory.
This includes states where the two $C=-$ glueballs are arbitrarily far apart.
It is interesting to understand how this can be consistent with the apparent absence of
single-particle $C=-$ states. This is a question that turns out to have a simple
solution, as we have shown very recently in 
\cite{MT_Pf},
although it is not clear at this stage whether the methods developed in that paper
will be useful in helping to resolve the puzzle that is the topic of the present paper.

In the next Section we briefly describe our lattice setup. In Section~\ref{section_strategy}
we explain in more detail how pairs of (anti)fundamental flux tubes affect the small
volume glueball spectra of $SU(N)$ gauge theories, and we provide some explicit calculations,
in both $SU(2)$ and $SU(4)$, that provide evidence for this. We then
move on, in Section~\ref{section_son}, to provide evidence for exactly the same
behaviour in $SO(3)$, $SO(4)$ and $SO(6)$ gauge theories. Section~\ref{section_string} 
is something of an aside: we provide some evidence that the fundamental flux tube in
$SO(6)$ is a bound state, presumably of two spinorial flux tubes. We then finish with some
conclusions.

\section{Lattice preliminaries}
\label{section_lattice}

Our lattice calculations are standard and we refer to
\cite{AAMT_SUN}
for a detailed description of the $SU(N)$ calculations, and to
\cite{RLMT_GK}
for details of our $SO(N)$ calculations. We simply remark here that
for $SO(N)$ our degrees of freedom are $N\times N$ real orthogonal matrices
$U_l$ with unit determinant, assigned to the links $l$ of the cubic periodic
space-time lattice, while for $SU(N)$ the matrices are $N\times N$ complex
unitary matrices with unit determinant. The partition function is 
\begin{equation}
Z=\int {\cal{D}}U \exp\{-\beta S[U]\}
\label{eqn_Z}
\end{equation}
where ${\cal{D}}U$ is the Haar measure and we use the standard plaquette action
for $S[U]$, and  $\beta=2N/ag^2$ where $ag^2$ is the dimensionless running
coupling on the length scale of the lattice spacing $a$. (We recall that
$g^2$ has dimensions $[m]^1$ in $D=2+1$.)

Ground state masses $M$ are calculated from the asymptotic time
dependence of correlators, i.e.
\begin{equation}
\langle \phi^\dagger(t) \phi(0) \rangle 
= \sum_n |\langle n|\phi|vac \rangle|^2 e^{-E_n t} 
\stackrel{t\to\infty}{\propto} e^{-Mt} 
\label{eqn_M}
\end{equation}
where $M$ is the mass of the lightest state with the quantum numbers
of the operator $\phi$. The operator $\phi$ will be the product of
link matrices around some closed path, with the trace then 
taken. To calculate the excited states $E_n$ in eqn(\ref{eqn_M}), one
calculates (cross)correlators of several operators and uses these as a
basis for a systematic variational calculation in $e^{-Ht_0}$ where $H$
is the Hamiltonian (corresponding to our lattice transfer matrix) and
$t_0$ is some convenient distance. (Typically we choose $t_0=a$.) 
To have good overlaps onto the
desired states, so that one can evaluate masses at values of $t$
where the signal has not yet disappeared into the statistical noise,
one uses blocked and smeared operators. (For more details see e.g.
\cite{AAMT_SUN,RLMT_GK}.)
For $SO(N)$ at the smallest values of $N$, these operators, as used in
\cite{RLMT_GK},
do not have very good overlaps onto the desired states. In
\cite{MT_GK}
we have improved the algorithm for both $SO(3)$ and $SO(4)$ and
are confident that our mass estimates in the present paper are reliable.

We label glueball states by spin $J$, parity $P$, and charge conjugation $C$.
Of course on our square lattice this is misleading since many continuum
states fall into the same representation, e.g. $J=0,4,8,..$ are
all in the same representation of the lattice rotation group. 
We shall take the convention of labelling the lightest
state by the smallest spin $J$ in that representation. This will be
correct for the lightest $0^+$ and $2^+$ states, the states of interest
in this paper, but it would not be correct, for example, for the
lightest `$0^-$' which is known to be in fact a $4^-$
\cite{HMMT_J}.

Since we are interested in the qualitative behaviour of glueball masses as the spatial
volume decreases, rather than in some precision calculation of mass ratios,  
we do not attempt an extrapolation to the continuum limit, but rather perform
calculations at a single bare coupling (i.e at a single lattice spacing) that is chosen
to be small enough that lattice corrections should be insignificant. Our specific
choices are $\beta=12$ for $SU(2)$, $\beta=13.7$ for $SO(4)$, $\beta=7$ for $SO(3)$,
$\beta=46$ for $SO(6)$, and $\beta=63$ for $SU(4)$. For $SU(2)$ and $SU(4)$ the
lattice corrections to $m_G/\surd\sigma$ are much less than $1\%$ (see Tables 8,10 of
\cite{AAMT_SUN})
For $SO(4)$ and $SO(6)$ the lattice corrections are also smaller than $1\%$
(see Tables 12,14,21,23, 28 of
\cite{RLMT_GK}).
This also appears to be the case for $SO(3)$ (see Tables 11,20,28 of
\cite{RLMT_GK})
but the instability of the flux tube means that the errors are relatively large.
In addition $\beta=7$ is on the weak edge of the weak-strong coupling transition
but this appears to have no effect on the states relevant to our calculations.
In summary, we believe that our lattice calculations provide an accurate
picture of what occurs in the corresponding continuum limits.

At various points in this paper we will make a statement that some mass $\mu$ in
an $SO(N)$ gauge theory equals a corresponding mass in the $SU(N^\prime)$ gauge
theory, where the groups share the same Lie algebra. What we mean by this is
that when one calculates the continuum limits of $\mu/g^2$ in $SO(N)$ and
in  $SU(N^\prime)$ then they are equal within errors once one relates the
two values of $g^2$ by the theoretically predicted factors
\cite{RLMT_GK}.

\section{Strategy: glueballs in small volumes}
\label{section_strategy}

As remarked in the Introduction, on small enough spatial volumes the energy of a
conjugate pair of winding flux tubes may become small enough for this state to become
the lightest glueball state. In particular this is most likely to occur for glueball
quantum numbers that are accessible to a pair of flux tubes that are in their ground state,
i.e. for $J^P=0^+,2^+$.  For other quantum numbers at least one flux tube has to be in an
excited state, which means its energy is considerably higher and so it may be that
the flux tube state never becomes lighter than the `normal' glueball. And clearly
all this is only relevant if the volume on which the glueball ground state changes
character from a normal glueball to a pair of winding flux tubes is large enough
to still support a confining vacuum.

Of course on a very small volume a conjugate pair of flux tubes will strongly overlap
and interact, and this may increase the energy of the flux tube state sufficiently that
it never becomes light enough to take over as the lightest glueball. So we can be 
confident that a crossing will occur if the lightest glueball is sufficiently
heavy that the level crossing takes place on a large enough volume for the flux
tube interaction to be modest. Since the $J^P=2^+$ glueball is heavier than
the $0^+$, most of our calculations will focus on that state.

Much of the past discussion of this particular small volume physics has been in the
context of $D=3+1$ $SU(N)$ gauge theories. Some early references are
\cite{CM_V}
and a more recent reference is
\cite{HM_V}.
We shall now examine how this works in $D=2+1$ for our $SU(2)$ and $SU(4)$ gauge
theories.  

\subsection{SU(2)}
\label{subsection_su2}

We shall illustrate how this works in the $D=2+1$  $SU(2)$ gauge theory working at
$\beta=4/ag^2=12.0$. We calculate the lightest $2^+$, $2^-$ and $0^+$ glueball masses
on a range of lattice space-time volumes, $l_x\times l_y\times l_t$,  as listed in
Table~\ref{table_Vsu2}. Here we use $l_x=l_y=l$ so as to maintain the square lattice
rotation symmetry that allows us to identify $J=0$ and $J=2$ states.
We also calculate the energy $aE_f(l_x)$ of the lightest flux tube that winds around the
periodic $x$-direction. We list twice that value in  Table~\ref{table_Vsu2} as that
provides us an estimate of the energy of the state containing a flux tube and its conjugate,
although, since it neglects the interaction energy between the two flux tubes, it can only
serve as a rough estimate, especially on smaller volumes.

We display how the lightest $SU(2)$ glueball mass varies with the spatial volume $V=l^2$
in Fig.\ref{fig_Vsu2}. We have chosen to express everything in units of the $V=\infty$ mass
gap, i.e. the mass of the lightest $0^+$ glueball on large volumes. In the continuum limit
and on an infinite volume the $2^+$ and $2^-$ states should be degenerate. (Parity
doubling for $J\neq 0$ in 2 space dimensions.) So as we decrease the volume, the
breaking of the near-degeneracy is a useful indication of the onset of important finite
volume effects. As we see in  Fig.\ref{fig_Vsu2} the $2^+$ and $2^-$ glueballs are
indeed (nearly) degenerate on our largest volumes. As we decrease $l$ we observe a
sudden breaking of the degeneracy close to the value of $l$ where the energy $2aE_f(l)$
of two flux tubes drops below the large-$V$ mass of the lightest $2^+$ glueball. As we
decrease $l$ further the lightest $2^+$ mass decreases approximately in step with the value
of  $2aE_f(l)$. Moreover on these smaller volumes the first excited $2^+$ state, which
on larger volumes was nearly degenerate with the first excited $2^-$, decomes nearly
degenerate with the lightest $2^-$. All this provides strong evidence for the
scenario of level crossing, where the $2^+$ ground state becomes, at small $l$, a state
of two winding flux tubes, while the first excited $2^+$ state is now the `normal' glueball
state that satisfies approximate parity doubling.

\begin{figure}[htb]
\begin	{center}
\leavevmode
\input	{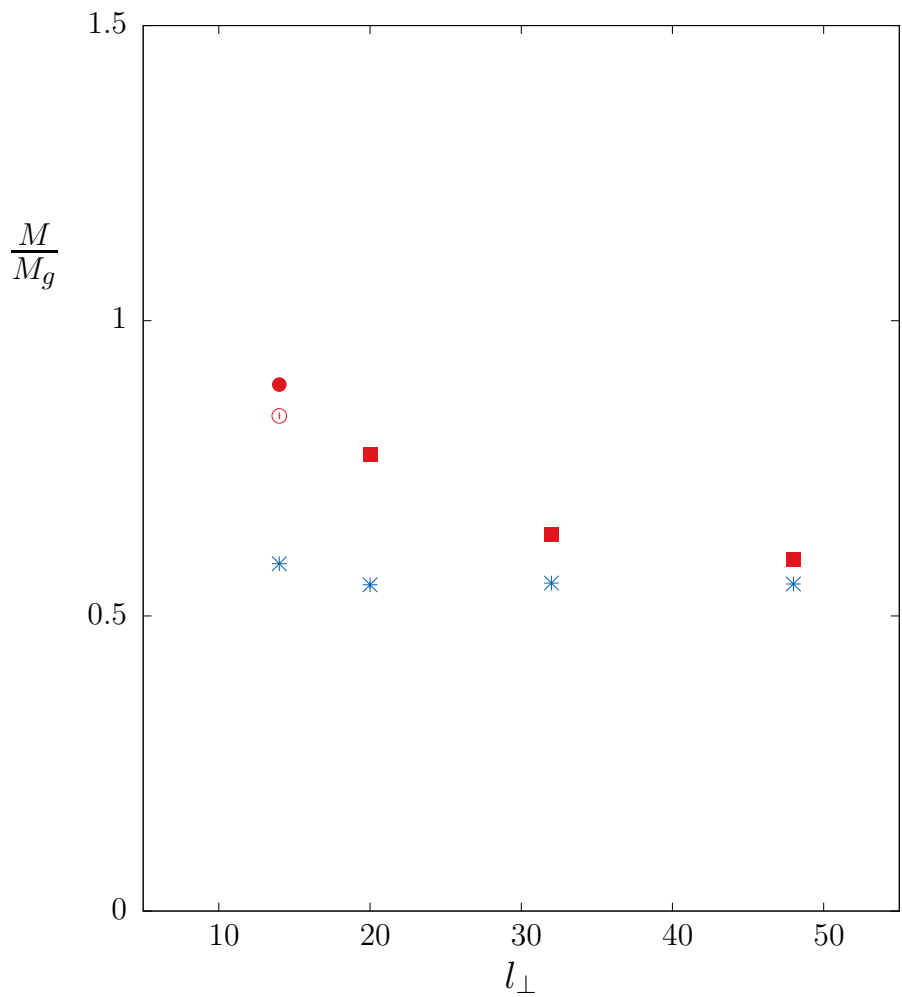}
\end	{center}
\caption{At $\beta=12.0$ in $SU(2)$ for our shortest flux
  tube, $l=14$: variation of lightest masses as a function of the transverse
  spatial size, $l_\perp$, in units of the infinite volume mass gap $M_g$.
  For $l_\perp=14$ the lightest $J^P=0^+$ ($\bullet$) and $J^P=2^+$ ($\circ$)
  masses are shown, and for $l_\perp > 14$ we show the mass gap ($\blacksquare$).
  Also displayed is twice the energy of the winding flux loop ($\star$).}
\label{fig_Vsu2perp}
\end{figure}

We also see from  Fig.\ref{fig_Vsu2} that as we decrease $l$ there is a slowly growing gap
between the $2^+$ ground state and the energy of two non-interacting flux tubes, $2aE_f(l)$.
This is presumably due to the interaction energy of the two flux tubes which will
increasingly overlap as $l$ decreases. (We also see that because the $0^+$ is much lighter,
any level crossing occurs at a much smaller value of $l$ so its interpretation as a
pair of flux tubes is much less compelling.) 
We would expect that if we increase the transverse size of the volume, i.e. 
$V = l\times l \to l\times l_\perp$ with $l_\perp \uparrow$, then the interaction energy
between the flux tubes will decrease because the physical overlap between the flux tubes 
decreases. Of course once $l\neq l_\perp$ we lose the $\pi/2$ rotation symmetry and
the would-be $0^+$ and $2^+$ operators will mix, so it only makes sense to talk of the
mass gap. This is what we plot in Fig.\ref{fig_Vsu2perp} for our smallest value $l=14$,
using the values listed in Table~\ref{table_Vsu2perp} (which have been obtained using
`rotationally symmetric' operators). We also list there 
the values of the lightest $0^+$ and $2^+$ glueballs on the symmetric
$l^2$ volume. We see how the mass gap appears to approach the energy of two non-interacting
flux tubes as $l_\perp/l$ increases, further confirming the above scenario.

We have not attempted a more systematic study because this interpretation of the small
volume behaviour of the $2^+$ and $0^+$ is not controversial. It would of course be
useful to show how the energy of the two interacting flux tubes varies with
the spatial volume but this would be a delicate calculation given the mixing of such
a state with the normal lightest glueballs and would go beyond the scope of this 
paper. We did however calculate the correlators of a few operators of the form
$l_f(p=0)\times l^\dagger_f(p=0)$ where $l_f(p=0)$ is (the trace of) a  winding
flux tube operator that is summed over all spatial sites, and hence has $p=0$.
This gives us an operator that, neglecting interactions, projects onto a state
with two conjugate $p=0$ winding flux tubes, i.e. with zero total and
relative momentum. Adding or subtracting such product operators winding around
the $x$ and $y$ tori gives us $J=0$ and $J=2$ operators. Of course what we are
interested in are the states of two flux tube states that interact and once one has
significant interactions (as we certainly will have in $SU(2)$)
our operator will also project onto pairs of flux tube states
with $p\neq 0$ (although the total momentum of the state must remain zero) as well
as mixing with `normal' glueballs. As the volume increases the allowed $p\neq 0$
values become smaller so that the energy spectrum becomes denser and it
becomes increasingly difficult to extract, accurately, the lightest flux tube
state. Nonetheless we find that on our smaller volumes we are more-or-less able
to do so and we list in Table~\ref{table_Vsu2ll} our best estimates of the
energies of the lightest $J=0$ and $J=2$  states composed of a pair of
conjugate flux tubes. (For $J=0$ we perform a vacuum subtraction, since otherwise
this would be the lightest state.) We see that as the volume increases these
energies are consistent with approaching those of two non-interacting flux tubes
and, indeed, a comparison with the $0^+$ and $2^+$ glueball masses in
Table~\ref{table_Vsu2} reinforces that interpretation.

\subsection{SU(4)}
\label{subsection_su4}

We recall that the operators that project onto the lightest glueball in a large
volume are formed by taking single traces of the path ordered product of
link matrices around contractible closed  loops and these are the operators
we use in our glueball correlators on all our volumes. The states composed of two
(conjugate) winding flux tubes correspond to double trace operators (up to the
effects of mixing). Thus their appearance in
glueball correlators will be increasingly suppressed as $N$ increases,
both for $SU(N)$ and for $SO(N)$, by the usual large-$N$ counting rules
\cite{large_N}.
It is thus necessary to check whether the small volume glueball dynamics described
above manifests itself as clearly in our $SU(4)$ glueball correlators as it
did in our above $SU(2)$ calculations.

Our $SU(4)$ calculations are performed at a bare coupling $\beta=8/ag^2=63.0$.
We list in Table~\ref{table_Vsu4} the lightest $2^+$, $2^-$ and $0^+$ masses
that we obtain from our glueball correlators, on the various spatial volumes
listed. We also list twice the energy, $aE_f(l)$, of the ground state of the
winding flux tube of length $l$ that carries fundamental flux. Since it will
be useful later on, we also list the corresponding energy $aE_{2A}(l)$ of a flux
tube carrying flux in the $k=2$ antisymmetric representation.

We display in Fig.\ref{fig_Vsu4} the $SU(4)$ analogue of our $SU(2)$ plot
in  Fig.\ref{fig_Vsu2}. The main features are similar in both plots: in
$SU(4)$ we have clear evidence, just as in $SU(2)$, for a level crossing as the
spatial volume decreases below the point where a pair of flux tubes becomes
lighter than the large-volume normal glueball mass. So as the volume decreases
beyond this value, one expects the corresponding glueball ground state to be
composed of a pair of conjugate winding flux tubes.

At a more detailed level there are some differences between Fig.\ref{fig_Vsu4}
and  Fig.\ref{fig_Vsu2}. In particular on the smallest volumes we see in
Fig.\ref{fig_Vsu4} that there is little difference between the value
of $2aE_f(l)$ and either the $0^+$ or the $2^+$ glueball mass. This is
in contrast to what we see in  Fig.\ref{fig_Vsu2}. The natural interpretation
is that what we are seeing is the suppression of the interaction
between the two flux tubes as $N$
increases, since standard counting tells us that the interaction energy between two
colour singlets vanishes as $N\to\infty$. Of course the same counting
also tells us that the overlap of the state consisting of two flux tubes onto the
single trace glueball operators should decrease as $N$ increases and so
the worry is that we could lose the corresponding signal in the statistical
noise of our glueball correlators. Although the overlaps do indeed appear to
decrease as we move from $SU(2)$ to $SU(4)$ we find, fortunately, that this does
not provide a significant obstacle to seeing this physics in $SU(4)$. This provides
us with some confidence that we can perform a useful comparison between $SU(4)$
and $SO(6)$.

\section{Glueballs in small volumes: SO(3), SO(4) and SO(6)}
\label{section_son}
\vspace*{0.2cm}

\subsection{SO(3) and SU(2)}
\label{subsection_so3}
\vspace*{0.2cm}

Since $SO(3)$ has the same Lie algebra as $SU(2)$ it is natural
to ask if the light glueball spectra are the same
\cite{RLMT_GK}.
In particular, if the ground states are the same in a finite volume
then we can infer that on small volumes the state is composed of
a pair of conjugate winding flux tubes that carry a flux which corresponds
to the fundamental of $SU(2)$, i.e. the spinorial of $SO(3)$.

So in Fig.\ref{fig_Vso3su2} we plot the lightest $SO(3)$ $0^+$ and $2^+$ glueball
masses (listed in Table~\ref{table_Vso3}) as a function of the spatial volume $V=l^2$.
The calculation is at an
inverse bare coupling $\beta=6/ag^2=7.0$ and we express all the quantities
in units of the large volume mass gap, $M_g$, at this coupling. We also plot
the corresponding quantities obtained in $SU(2)$ at $\beta=8/ag^2=12.0$,
together with twice the energy of the lightest flux tube, $2E_f(l)$, in $SU(2)$.
The $SU(2)$ quantities are also expressed in units of the large volume mass gap,
but this time that of $SU(2)$ rather than $SO(3)$. Since one finds that
in the continuum limit the large
volume $SU(2)$ and $SO(3)$ mass gaps are the same within small errors, we can
regard the $SU(2)$ and $SO(3)$ quantities in  Fig.\ref{fig_Vso3su2} as
being expressed with a common scale.

What we see in Fig.\ref{fig_Vso3su2} is that within quite small errors
the volume dependences of the $2^+$ and $0^+$  glueball masses are identical
in $SO(3)$ and $SU(2)$. So we can infer that on the smallest volumes the $2^+$
glueball ground state in $SO(3)$ is in fact a state composed of a
conjugate pair of spinorial flux tubes, corresponding to a conjugate pair
of fundamental flux tubes in $SU(2)$.

As in $SU(2)$ the $SO(3)$ glueball masses on the smallest volumes are
significantly heavier than $2E_f(l)$, the energy of two non-interacting
flux tubes, and just as for $SU(2)$ we interpret this gap
as being due to the interaction energy arising from the fact that on these small
volumes the flux tubes are necessarily overlapping. So we would expect that if we
perform calculations on an asymmetric lattice volume, $V=ll_\perp$,
and increase $l_\perp$ 
then the flux tubes will overlap less and this gap will decrease, just as
we observed for $SU(2)$ in Fig.\ref{fig_Vsu2perp}. We have
performed such a study for $SO(3)$ leading to the values listed
in Table~\ref{table_Vso3perp}, and we observe in  Fig.\ref{fig_Vso3perp}
the same behaviour as in $SU(2)$. All this helps to corroborate our interpretation
of the nature of the finite volume glueball ground states in the $SO(3)$ 
gauge theory.

\subsection{SO(4) and SU(2)}
\label{subsection_so4}
\vspace*{0.2cm}

The groups $SO(4)$ and $SU(2)\times SU(2)$ share the same Lie algebra so the naive
expectation is that the spectrum of  $SO(4)$ will be that of two mutually
non-interacting $SU(2)$ groups.
If this is so, then we expect the lightest $0^+$ and $2^+$ glueballs to have the
same masses in $SU(2)$ and in $SO(4)$, and on large volumes that indeed appears
to be the case
\cite{RLMT_GK}.
Since in $SO(4)$ the confining flux tube in the fundamental representation has a
string tension that is twice the fundamental $SU(2)$ string tension
\cite{RLMT_GK}
it will be approximately twice as massive for a given length $l$ of the flux tube.
So as we decrease $l$ the location of the level crossing where the lightest
glueball becomes a (conjugate) pair of these
flux tubes, and where its mass begins to decrease strongly with decreasing $l$,
should occur at a much smaller value of $l$ than in the case of $SU(2)$. If on the
other hand we find that this decrease in the glueball mass begins at the same value of
$l$ as in $SU(2)$, which would of course be consistent with the very naive expectation
that the $SO(4)$ and $SU(2)$ masses are the same at any volume, then this is evidence
for a level crossing with a (conjugate) pair of spinorial flux tubes (corresponding to
a pair of fundamentals in $SU(2)$).

\begin{figure}[htb]
\begin	{center}
\leavevmode
\input	{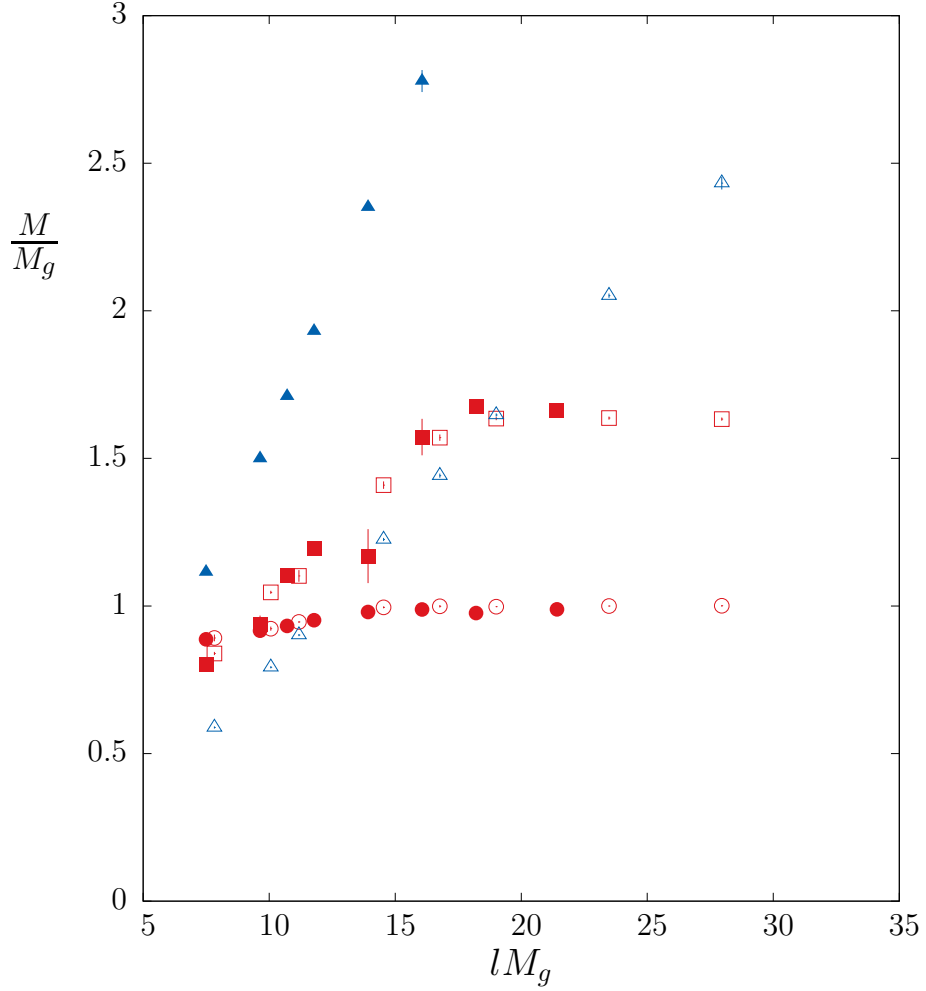}
\end	{center}
\caption{Lightest scalar ($\bullet$) and tensor ($\blacksquare$) glueballs
in SO(4) on various spatial volumes, $l^2$, all in units of the infinite 
volume mass gap $M_g$. And the same for SU(2) ($\circ,\square$ respectively).
Also shown is twice the fundamental flux loop mass in SU(2) ($\triangle$)
and SO(4) ($\blacktriangle$).}
\label{fig_Vso4su2}
\end{figure}

In Fig.\ref{fig_Vso4su2} we display how the masses of the lightest $0^+$ and $2^+$
glueballs (listed in Table~\ref{table_Vso4}) vary with the spatial volume
in $SO(4)$ at $\beta=13.7$ and in $SU(2)$
at $\beta=12.0$ using the corresponding $V=\infty$ mass gaps to set the scale.
(These mass gaps are equal in the continuum limit within small uncertainties
\cite{RLMT_GK}.)
We also show the energy of a pair of non-interacting fundamental flux tubes
that wind around the torus, for both $SU(2)$ and $SO(4)$. We see that the
latter energy is about twice the former at a given $l$, reflecting the
factor of two difference in the string tensions. We also see that the $2^+$
masses are essentially identical for the two groups, and that they start
decreasing rapidly at the same value of $l$. At this value of $l$ the energy
of two $SO(4)$ fundamental flux tubes is much too large for it to be driving
this decrease. So just as for $SO(3)$ we conclude that the $SO(4)$ gauge theory
contains states that are composed of a pair flux tubes that carry a flux that
corresponds to the fundamental of $SU(2)$. This despite the fact that we cannot
construct a wave-functional for a single such flux tube, using our $SO(4)$ gauge
fields, and so might naively expect that a single flux tube of this type would not
appear in the spectrum of the theory.

\subsection{SO(6) and SU(4)}
\label{subsection_so6}

We now turn to $SU(4)$ and $SO(6)$ which also share a common Lie algebra.
The fundamental of $SO(6)$ corresponds to the $k=2A$ of $SU(4)$, i.e the
totally antisymmetric piece of $f\otimes f$ of $SU(4)$, while the fundamental
$f$ of  $SU(4)$ corresponds to the spinorial of  $SO(6)$. Thus the relation
between the string tensions will be
\cite{AAMT_SUN}
\begin{equation}
  \frac{\sigma_{f,so6}}{\sigma_{sp,so6}}
  =
  \frac{\sigma_{2A,su4}}{\sigma_{f,su4}}
  \simeq 1.357(3).
\label{eqn_fso6fsu4}
\end{equation}
using an obvious notation. So the difference in energy between a pair of flux
tubes in the fundamental and spinorial representations will be much less
in  $SO(6)$ than in $SO(4)$. This would suggests that our evidence for spinorials might
be  less impressive for  $SO(6)$ than for $SO(3)$ or for $SO(4)$. In addition the
larger value of $N$ will reduce the overlap of our single trace operators
onto the double trace operators that are natural for two flux tubes
and might make it hard to see any level crossing at all.  However in
practice, as we have already seen for $SU(4)$ in Fig.\ref{fig_Vsu4},
the same large $N$ effects appear to reduce the interaction energy in
a way that makes the level crossing more stark, and this appears
to largely compensate for the negative factors decribed above.

\begin{figure}[htb]
\begin	{center}
\leavevmode
\input	{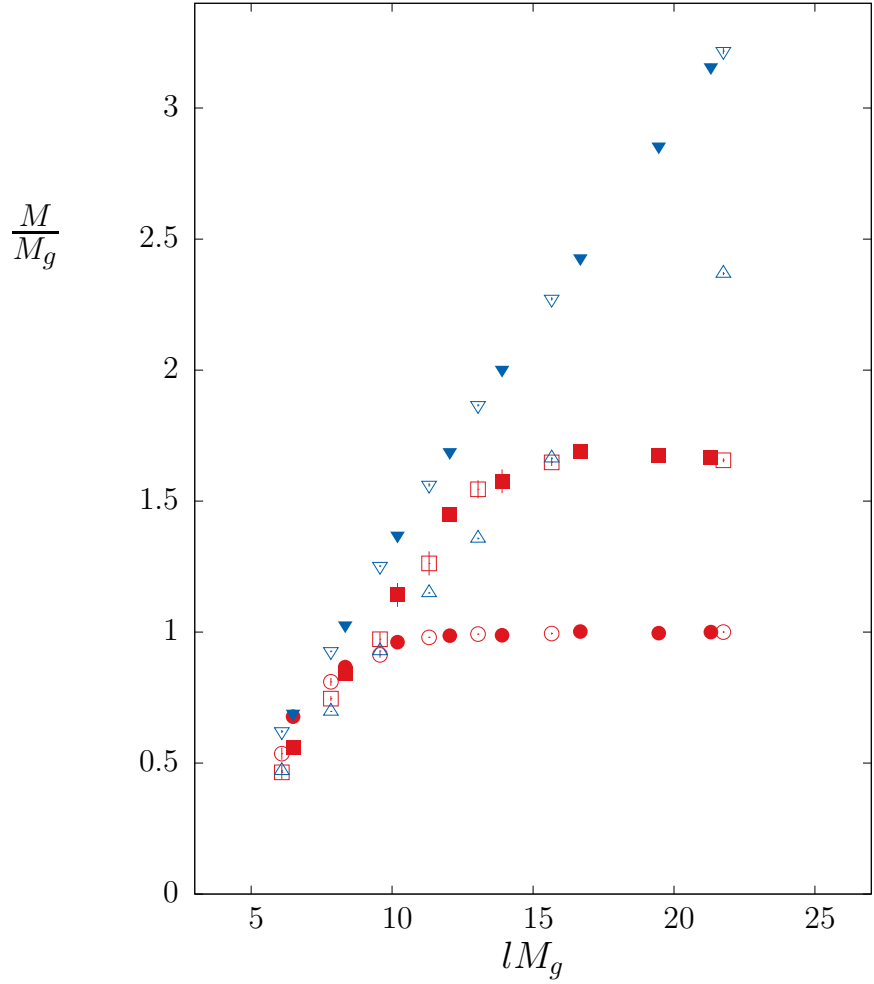}
\end	{center}
\caption{Lightest scalar ($\circ$) and tensor ($\Box$) glueballs
in SU(4) on various spatial volumes, $l^2$, all in units of the infinite 
volume mass gap $M_g$. And same for SO(6) ($\bullet,\blacksquare$ respectively).
Also shown is twice the fundamental flux loop energy in SU(4) ($\triangle$)
and in SO(6) ($\blacktriangledown$), and the $k=2$ flux loop energy in
SU(4) ($\triangledown$).}
\label{fig_Vso6su4}
\end{figure}

Turning now to our results, listed in Table~\ref{table_Vso6},
we show in Fig.\ref{fig_Vso6su4} the finite volume
dependence of our lightest $2^+$ and $0^+$ glueball masses calculated in
$SO(6)$ at $\beta=46$ and compare to that of $SU(4)$ calculated
at $\beta=63$. We also show the energy of two non-interacting
winding flux tubes for the case where both are in the fundamental of $SO(6)$
and also the case where both are in the fundamental of $SU(4)$ (corresponding
to the spinorial of $SO(6)$) and in the $k=2A$ of $SU(4)$ (corresponding
to the fundamental of $SO(6)$). As before we use the infinite volume
mass gaps as our scale, noting that we know these
to be equal in the continuum limit within small uncertainties
\cite{RLMT_GK}.
We observe in  Fig.\ref{fig_Vso6su4} that the energy of two free flux tubes
in the fundamental of $SO(6)$ coincides with that of flux tubes in the
$k=2A$ of $SU(4)$ over the whole range of volumes, indicating that the differing
global properties of the  $SO(6)$ and  $SU(4)$ groups are unimportant
for this physics. We also observe that the variation with volume of the
lightest $2^+$ and $0^+$ glueball masses appears to be identical for
$SO(6)$ and  $SU(4)$. In
particular the onset of the rapid decrease in the masses with decreasing $l$
appears to occur at the same value of $l$ for $SO(6)$ and $SU(4)$, indicating
an identical level crossing dynamics. So it is very plausible that the
small volume $SO(6)$ glueballs are composed of a pair of winding (conjugate)
spinorial flux tubes. Note that the comparison with $SU(4)$ is crucial
here: if we were relying only on the $SO(6)$ results, then it would
be difficult to argue from Fig.\ref{fig_Vso6su4} that the level crossing
is associated with pairs of spinorial rather than fundamental flux tubes,
because these do not differ very much in energy.

\section{SO(6) flux tube as a bound state}
\label{section_string}

In $SU(4)$ one finds
\cite{AAMT_string}
that the gap between the ground and first excited states of the $k=2A$
winding flux tube is roughly constant with $l$ as long as $l$ is small
enough that the higher excited states are significantly heavier (to avoid
ambiguities due to level crossings). This is in marked
contrast to the fundamental flux tube where the first excited state
closely tracks the `Nambu-Goto' string formula even down to very small values
of $l$
\cite{AAMT_string}.
A natural explanation of this difference is that the $k=2A$
spectrum is more complicated because the $k=2A$ flux tube should be
regarded as a (loosely) bound state of two fundamental flux tubes, with
the binding being encoded in the world sheet action by a massive particle
which will then appear in the flux tube spectrum
\cite{AAMT_string}.
(For a relevant discussion of the world sheet action see
\cite{worldsheet}
and references therein.) Then one
interprets this first excited state as being composed of the
ground state flux tube with this massive excitation bound to it
\cite{AAMT_string}.
Conversely the existence of this `anomalous' excitation encourages
the interpretation of the $k=2A$ flux tube as a (loosely) bound
state of two fundamental flux tubes. It is thus interesting to
see if the $SO(6)$ fundamental flux tube spectrum also shows
such an `anomalous' excited state. If it does then we would be motivated
to interpret the  $SO(6)$ fundamental flux tube as a loosely bound state of
two spinorial flux tubes.

\begin{figure}[htb]
\begin	{center}
\leavevmode
\input	{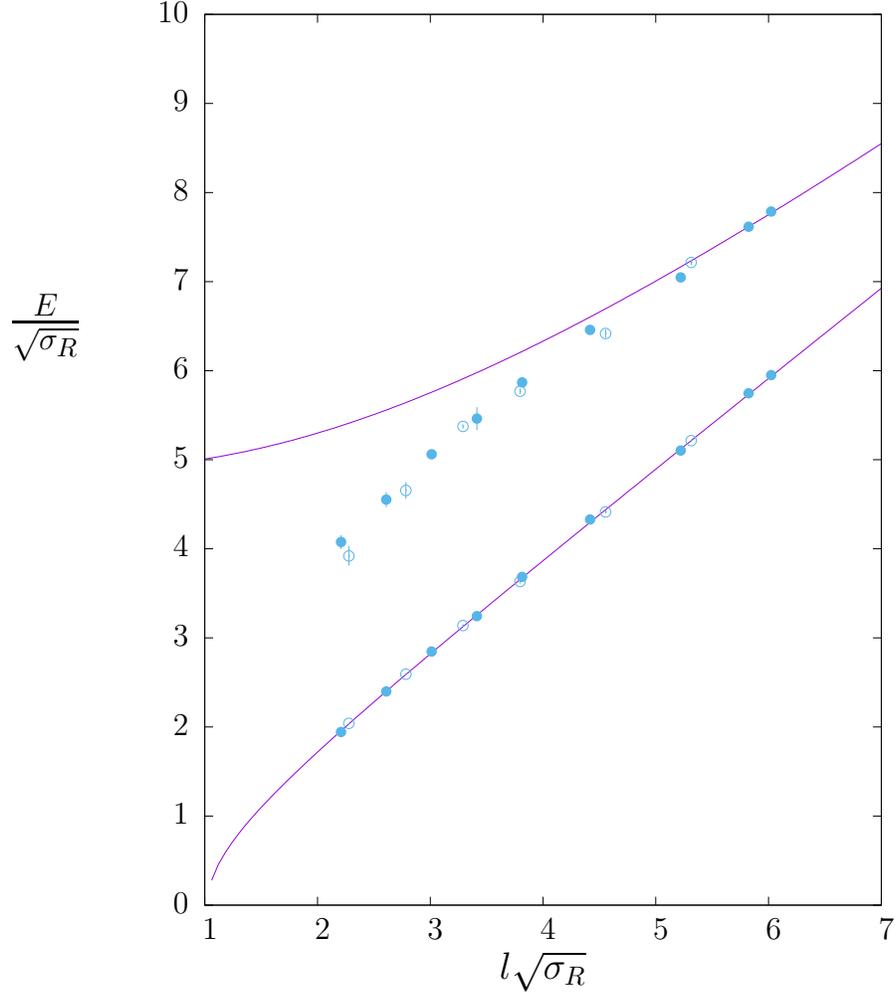}
\end	{center}
\caption{Energies of the lightest two winding flux tubes of length $l$ on spatial
  volumes $l\times l_\perp$ with $l_\perp \gg l$ for small $l$. For $SU(4)$ ($\bullet$)
  and $SO(6)$ ($\circ$), scaled by $\sigma_R=\sigma_{2A}$ and $\sigma_R=\sigma_f$
  respectively. Lines are `Nambu-Goto' predictions.}
\label{fig_DElso6su4}
\end{figure}

So in Fig.\ref{fig_DElso6su4} we plot the energies of the ground
and first excited winding flux tubes in $SO(6)$. For comparison
we also plot the corresponding energies of the $k=2A$ flux tubes
in $SU(4)$. The energies and lengths are expressed in units of the
corresponding string tensions, which are known to be equal
(within small uncertainties) in the continuum limit
\cite{RLMT_GK}.
We see that the $SU(4)$ and $SO(6)$ spectra are identical within
small errors. That is to say, we have some evidence that the
fundamental flux tube of $SO(6)$ is in fact a `loosely' bound
state of two spinorial flux tubes.

\section{Conclusions}
\label{section_concl}

By calculating in $SO(3)$, $SO(4)$ and $SO(6)$ gauge theories the finite
volume behaviour of the lightest $2^+$ glueball and, to a lesser extent,
that of the $0^+$, and by comparing to the corresponding $SU(N)$ gauge
theories with the same Lie algebra, we obtained strong evidence
that on small volumes the lightest glueball becomes a state composed
of a pair of loosely bound spinorial flux tubes that wind around the
spatial volume.

This is interesting because in our $SO(N)$ gauge theories we cannot
construct from the $SO(N)$ lattice fields operators that project onto
single spinorial flux tubes and so one might naively expect that these
would not exist in the spectrum of the theory. Of course our small
volume glueball states are typically composed of a pair of winding 
(conjugate) spinorial flux tubes that are physically overlapping
and which are interacting rather than free. However, as we have
seen, if we work on asymmetric $l\times l_\perp$ spatial volumes and
increase $l_\perp$ then this interaction energy appears to decrease towards
zero. Thus the theory does seem to include states that are composed of
a pair of winding spinorial flux tubes that are arbitrarily
far apart. Having established this brings into sharp focus the
interesting question of how to square
it with the naive expectation that the theory does not include states
composed of a single spinorial flux tube,

\section*{Acknowledgements}

The numerical computations were carried out on the computing cluster
in Oxford Theoretical Physics. The author is grateful to both Oxford
Theoretical Physics and to All Souls College for their support of this
research.

%
%
%
%


\newpage

\begin{table}[htb]
\begin{center}
\begin{tabular}{|c|c|c|c|c|}\hline
\multicolumn{5}{|c|}{ $SU(2) \quad \beta=12.0$ } \\ \hline
$l_x\times l_y\times l_t$ & $am_{2^+}$ & $am_{2^-}$ & $am_{0^+}$ & $2aE_f(l_x)$   \\ \hline
$50\times 50\times 40$  & 0.9132(25)  & 0.9207(28)  & 0.5594(14) &  1.360(12)  \\
$42\times 42\times 40$  & 0.9146(20)  & 0.9155(24)  & 0.5589(10) &  1.1470(34) \\
$34\times 34\times 40$  & 0.9142(23)  & 0.9165(25)  & 0.5578(9)  &  0.9206(28) \\
$30\times 30\times 48$  & 0.8776(59)  & 0.9215(30)  & 0.5585(21) &  0.8062(32) \\
$26\times 26\times 56$  & 0.7876(65)  & 0.9222(36)  & 0.5565(17) &  0.6852(28) \\
$20\times 20\times 56$  & 0.616(11)   & 0.9122(61)  & 0.5288(17) &  0.5040(18) \\
$18\times 18\times 68$  & 0.5852(24)  & 0.9177(71)  & 0.5161(39) &  0.4430(12)  \\
$14\times 14\times 80$  & 0.4690(32)  & 0.8893(85)  & 0.4986(65) &  0.3290(12)  \\ \hline
\end{tabular}
\caption{Lightest $J^P=2^+,2^-,0^+$ glueball masses on various
  $l_x\times l_y$ spatial lattice volumes. Also the energy $E_f(l_x)$
  of the lightest fundamental flux tube that winds around the
  $x$-torus. In $SU(2)$ at an inverse bare coupling
  $\beta = 4/ag^2 =12.0$.}
\label{table_Vsu2}
\end{center}
\end{table}

\begin{table}[htb]
\begin{center}
\begin{tabular}{|c|c|c|c|c|}\hline
\multicolumn{5}{|c|}{ $SU(2) \quad \beta=12.0$ } \\ \hline
$l\times l_\perp\times l_t$ & $am_{2^+}$ & $am_{0^+}$ & $am_g$ & $2aE_f(l)$   \\ \hline
$14\times 14\times 80$  & 0.4690(32)  & 0.4986(65) &    --        & 0.3290(12)  \\
$14\times 20\times 80$  &    --          &      --       &  0.4328(56)  & 0.3090(32)   \\
$14\times 32\times 80$  &    --          &      --       &  0.3565(65)  & 0.3106(14)   \\
$14\times 48\times 80$  &    --          &      --       &  0.3333(33)  & 0.3098(16)   \\ \hline
\end{tabular}
\caption{Lightest $J^P=2^+,0^+$ glueball masses and the mass gap $m_g$ on the 
  spatial lattice volumes shown. Also the energy $E_f(l_x)$
  of the lightest fundamental flux tube that winds around the
  $x$-torus. In $SU(2)$ at an inverse bare coupling
  $\beta = 4/ag^2 =12.0$.}
\label{table_Vsu2perp}
\end{center}
\end{table}

\begin{table}[htb]
\begin{center}
\begin{tabular}{|c|c|c|c|}\hline
\multicolumn{4}{|c|}{ $SU(2)$: \quad $l_f(p=0) l^\dagger_f(p=0)$ } \\ \hline
$l_x\times l_y$ & $am_{2^+}$ & $am_{0^+}$ & $2aE_f(l)$   \\ \hline
$14\times 14$  & 0.4670(30)  & 0.4945(13) & 0.3290(12)  \\
$18\times 18$  & 0.5822(25)  & 0.5432(43) & 0.4430(12)   \\
$20\times 20$  & 0.618(11)   & 0.5661(57) & 0.5040(18)   \\
$26\times 26$  & 0.626(105)  & 0.668(11)  & 0.6852(28)   \\  \hline
\end{tabular}
\caption{Energies of lightest $J^P=2^+,0^+$ states formed from the product
  of a pair of (conjugate) winding flux tube operators  on various 
  $l_x\times l_y$ spatial lattice volumes. Also twice the energy $E_f(l_x)$
  of the lightest fundamental flux tube that winds around the
  $x$-torus. In $SU(2)$ at an inverse bare coupling
  $\beta = 4/ag^2 =12.0$.}
\label{table_Vsu2ll}
\end{center}
\end{table}

\begin{table}[htb]
\begin{center}
\begin{tabular}{|c|c|c|c|c|c|}\hline
\multicolumn{6}{|c|}{ $SU(4) \quad \beta=63.0$ } \\ \hline
$l_x\times l_y\times l_t$ & $am_{2^+}$ & $am_{2^-}$ & $am_{0^+}$ & $2aE_f(l_x)$ & $2aE_{2A}(l_x)$   \\ \hline
$50\times 50\times 56$  & 0.7203(33) & 0.7219(41) &  0.4350(15) & 1.0302(26) & 1.3996(52)   \\
$36\times 36\times 56$  & 0.7167(60) & 0.7258(20) &  0.4326(18) & 0.7236(14) & 0.9882(28)    \\
$30\times 30\times 60$  & 0.672(15)  & 0.7293(22) &  0.4316(12) & 0.5904(18) & 0.8096(20)    \\
$26\times 26\times 70$  & 0.549(20)  & 0.695(15)  &  0.4262(12) & 0.5000(16) & 0.6792(30)      \\
$22\times 22\times 80$  & 0.423(12)  & 0.680(16)  &  0.3972(47) & 0.4034(12) & 0.5446(18)      \\
$18\times 18\times 90$  & 0.3244(42) & 0.525(13)  &  0.3524(67) & 0.3032(8)  & 0.4030(14)      \\  
$14\times 14\times 100$ & 0.202(11)  & 0.381(14)  &  0.233(10)  & 0.2044(12) & 0.2700(26)    \\  \hline
\end{tabular}
\caption{Lightest $J^P=2^+,2^-,0^+$ glueball masses on various
  $l_x\times l_y$ spatial lattice volumes. Also twice the energy 
  of the lightest fundamental, $E_f(l_x)$, and $k=2$, $E_{2A}(l_x)$, flux tubes
  that wind around the spatial torus. In $SU(4)$ at an inverse bare coupling
  $\beta = 8/ag^2 =63.0$.}
\label{table_Vsu4}
\end{center}
\end{table}

\begin{table}[htb]
\begin{center}
\begin{tabular}{|c|c|c|c|}\hline
\multicolumn{4}{|c|}{ $SO(3) \quad \beta=7.0$ } \\ \hline
$l_x\times l_y\times l_t$ & $am_{2^+}$ & $am_{2^-}$ & $am_{0^+}$   \\ \hline
$62\times 62\times 48$  & 0.6541(94)  & 0.661(5)   & 0.4004(21)  \\
$46\times 46\times 48$  & 0.6582(83)  & 0.661(8)   & 0.4003(17) \\
$38\times 38\times 48$  & 0.587(15)   & 0.665(5)   & 0.3961(20)  \\
$34\times 34\times 54$  & 0.513(11)   & 0.651(8)   & 0.3866(40)     \\
$30\times 30\times 60$  & 0.4704(51)  & 0.648(7)   & 0.3793(23)    \\
$26\times 26\times 72$  & 0.4160(33)  & 0.652(9)   & 0.3699(15)     \\
$22\times 22\times 80$  & 0.3653(23)  & 0.638(14)  & 0.3502(74)     \\
$18\times 18\times 100$ & 0.3051(39)  & 0.582(3)   & 0.3374(20)     \\ \hline
\end{tabular}
\caption{Lightest $J^P=2^+,2^-,0^+$ glueball masses on various
  $l_x\times l_y$ spatial lattice volumes.  In $SO(3)$ at an inverse
  bare coupling $\beta = 6/ag^2 =7.0$.}
\label{table_Vso3}
\end{center}
\end{table}

\begin{table}[htb]
\begin{center}
\begin{tabular}{|c|c|c|c|}\hline
\multicolumn{4}{|c|}{ $SO(3) \quad \beta=7.0$ } \\ \hline
$l\times l_\perp\times l_t$  & $am_{0^+}$  & $am_{2^+}$ & $am_{g}$   \\ \hline
$18\times 18\times 100$  & 0.3374(20)  & 0.3051(39) &  --            \\
$18\times 30\times 100$  &  --         &  --        & 0.2539(69)     \\
$18\times 60\times 100$  &  --         &  --        & 0.2165(31)     \\ \hline
\end{tabular}
\caption{Lightest $J^P=2^+,0^+$ glueball masses and the mass gap $m_g$ on various
  $l_x\times l_y$ spatial lattice volumes.  In $SO(3)$ at an inverse
  bare coupling $\beta = 6/ag^2 =7.0$.}
\label{table_Vso3perp}
\end{center}
\end{table}

\begin{table}[htb]
\begin{center}
\begin{tabular}{|c|c|c|c|c|}\hline
\multicolumn{5}{|c|}{ $SO(4) \quad \beta=13.7$ } \\ \hline
$l_x\times l_y\times l_t$ & $am_{2^+}$ & $am_{2^-}$ & $am_{0^+}$ & $2aE_f(l_x)$   \\ \hline
$40\times 40\times 44$  & 0.8908(82)  & 0.862(17)   & 0.5294(89)  & 2.016(28)  \\
$34\times 34\times 44$  & 0.8975(81)  & 0.9145(36)  & 0.5227(86)  & 1.745(15)  \\
$30\times 30\times 48$  & 0.842(33)   & 0.878(16)   & 0.5292(67)  & 1.488(20)  \\
$26\times 26\times 56$  & 0.626(49)   & 0.887(10)   & 0.5246(61)  & 1.259(5)  \\
$22\times 22\times 56$  & 0.640(12)   & 0.8889(50)  & 0.5096(27)  & 1.034(3)  \\
$20\times 20\times 56$  & 0.590(5)    & 0.8955(54)  & 0.4994(38)  & 0.9162(46)  \\
$18\times 18\times 56$  & 0.501(17)   & 0.8884(61)  & 0.4906(38)  & 0.8028(28)  \\
$14\times 14\times 68$  & 0.4288(89)  & 0.834(11)   & 0.4751(75)  & 0.5972(14)  \\ \hline
\end{tabular}
\caption{Lightest $J^P=2^+,2^-,0^+$ glueball masses on various
  $l_x\times l_y$ spatial lattice volumes. Also twice the energy $E_f(l_x)$
  of the lightest fundamental flux tube that winds around the
  $x$-torus. In $SO(4)$ at an inverse bare coupling
  $\beta = 8/ag^2 =13.7$.}
\label{table_Vso4}
\end{center}
\end{table}

\begin{table}[htb]
\begin{center}
\begin{tabular}{|c|c|c|c|c|}\hline
\multicolumn{5}{|c|}{ $SO(6) \quad \beta=46.0$ } \\ \hline
$l_x\times l_y\times l_t$ & $am_{2^+}$ & $am_{2^-}$ & $am_{0^+}$ & $2aE_f(l_x)$   \\ \hline
$46\times 46\times 48$  & 0.7718(51) & 0.7729(63) & 0.4633(24) &  1.4628(60)  \\
$42\times 42\times 44$  & 0.7753(52) & 0.7675)68) & 0.4614(15) &  1.3204(48)    \\
$36\times 36\times 44$  & 0.7832(23) & 0.7881(27) & 0.4643(18) &  1.1204(34)    \\
$30\times 30\times 50$  & 0.730(21)  & 0.760(12)  & 0.4577(25) &  0.9276(36)     \\
$26\times 26\times 56$  & 0.671(13)  & 0.780(5)   & 0.4569(24) &  0.7822(28)    \\
$22\times 22\times 60$  & 0.529(21)  & 0.768(6)   & 0.4455(15) &  0.6344(22)    \\
$18\times 18\times 90$  & 0.390(10)  & 0.706(12)  & 0.4014(43) &  0.4758(22)    \\ 
$14\times 14\times 100$ & 0.226(13)  & 0.470(11)  & 0.3137(49) &  0.3196(20)    \\ \hline
\end{tabular}
\caption{Lightest $J^P=2^+,2^-,0^+$ glueball masses on various
  $l_x\times l_y$ spatial lattice volumes. Also twice the energy $E_f(l_x)$
  of the lightest fundamental flux tube that winds around the
  $x$-torus. In $SO(6)$ at an inverse bare coupling
  $\beta = 12/ag^2 =46.0$.}
\label{table_Vso6}
\end{center}
\end{table}


\clearpage

\begin{figure}[htb]
\begin	{center}
\leavevmode
\input	{plot_Vsu2.tex}
\end	{center}
\caption{Lightest $J^P=0^+$ ($\bullet$), $2^+$ ($\blacksquare$) and  $2^-$
  ($\star$) glueball masses in SU(2) at $\beta=12.0$  on various spatial volumes, $l^2$,
  all in units of the infinite volume mass gap $M_g$. The $2^-$ has been shifted for clarity.
  Also shown is the first excited $2^+$ ($\square$) glueball mass and twice the fundamental
  flux loop energy ($\blacktriangle$).}
\label{fig_Vsu2}
\end{figure}

\begin{figure}[htb]
\begin	{center}
\leavevmode
\input	{plot_Vsu4.tex}
\end	{center}
\caption{Lightest $J^P=0^+$ ($\bullet$), $2^+$ ($\blacksquare$) and  $2^-$
  ($\star$) glueball masses in SU(4) at $\beta=63.0$  on various spatial volumes, $l^2$,
  all in units of the infinite volume mass gap $M_g$. The $2^-$ has been shifted for clarity.
  Also shown is the first excited $2^+$ ($\square$) glueball mass and twice the fundamental
  flux loop energy ($\blacktriangle$).}
\label{fig_Vsu4}
\end{figure}

\begin{figure}[htb]
\begin	{center}
\leavevmode
\input	{plot_Vso3su2.tex}
\end	{center}
\caption{Lightest scalar ($\bullet$) and tensor ($\blacksquare$) glueballs
in SO(3) on various spatial volumes, $l^2$, all in units of the infinite 
volume mass gap $M_g$. And same for SU(2) ($\circ,\Box$ respectively).
Also shown is twice the fundamental flux loop mass in SU(2) ($\triangle$).}
\label{fig_Vso3su2}
\end{figure}

\begin{figure}[htb]
\begin	{center}
\leavevmode
\input	{plot_Vso3perp.tex}
\end	{center}
\caption{At $\beta=7.0$ in $SO(3)$ for our shortest flux
  tube, $l=18$: variation of lightest masses as a function of the transverse
  spatial size, $l_\perp$, in units of the infinite volume mass gap $M_g$.
  For $l_\perp=18$ the lightest $J^P=0^+$ ($\bullet$) and $J^P=2^+$ ($\circ$)
  masses are shown, and for $l_\perp > 18$ we show the mass gap ($\blacksquare$).
  Also shown are estimates of twice the energy of the fundamental
  $SU(2)$ flux loop energy ($\star$).}
\label{fig_Vso3perp}
\end{figure}

\end{document}